\documentclass[twocolumn,pre,floatfix]{revtex4}
\usepackage{dcolumn}
\usepackage{graphicx}
\usepackage{graphicx,amssymb,amsmath}
\usepackage{natbib}
\begin{document}


\title{On the mobility of extended bodies in viscous films and membranes}
\author{Alex J.~Levine}
\affiliation{The Institute for Theoretical Physics, University of
California, Santa Barbara CA 93106} \affiliation{Department of
Physics, University of Massachusetts, Amherst Amherst MA 01060 USA.}
\email[]{levine@physics.umass.edu}

\author{T.B.~Liverpool}
\affiliation{The Institute for Theoretical Physics, University of
California, Santa Barbara CA 93106} \affiliation{Department of Applied Mathematics, University of Leeds, Leeds LS2 9JT, England. } \email[]{tannie@maths.leeds.ac.uk}

\author{F.C.\ MacKintosh}
\affiliation{The Institute for
Theoretical Physics, University of California, Santa Barbara CA 93106}
\affiliation{Division of Physics \& Astronomy, Vrije Universiteit
1081 HV Amsterdam, The Netherlands}
\email[]{fcm@nat.vu.nl}

\date{\today}
\begin{abstract}
We develop general methods to calculate the mobilities of extended
bodies in (or associated with) membranes and films. We demonstrate
a striking difference between in-plane motion of rod-like
inclusions and the corresponding case of bulk (three-dimensional)
fluids: for rotations and motion perpendicular to the rod axis, we
find purely local drag, in which the drag coefficient is purely
algebraic in the rod dimensions. These results, as well as the
calculational methods are applicable to such problems as the
diffusion of objects in or associated with Langmuir films and
lipid membranes. The methods can also be simply extended to treat
viscoelastic systems.
\end{abstract}

\maketitle

\section{Introduction}
\label{intro}

The motion of objects in biomembranes is important in many
cellular processes. These objects are in many cases extended,
macromolecular inclusions such as proteins \cite{Stein,Spooner} or
``rafts'' \cite{Simons,Pralle} of lipids. Thus, these can often be
viewed as macroscopic objects moving in a continuum fluid
environment. Studies of the motion of such inclusions in
amphaphilic films \cite{Steffen,Klingler} and cell membranes have
a long and interesting history. There are discrepancies in the
early literature on protein diffusion in cell membranes, for
instance, because of confusion over the applicability of three-
versus two-dimensional diffusion to this case
\cite{Saffman,Hughes}. One of the most important contributions in
this area was that of Saffman \cite{Saffman}, who noted that the
two-dimensional motion in real membranes induces flows in the
surrounding bulk (three-dimensional) fluid. He showed that the
linearized Stokes-law does not describe the motion of inclusions
in or bound to membranes. Rather, this represents a singular
perturbation problem, and the drag coefficient is not a linear
function of its size and the viscosity. In fact, the dependence is
logarithmic, and thus the mobilities and diffusion coefficients of
proteins in membranes are nearly independent of the object's size,
in practice.

Furthermore, there is a
length-scale determined by the ratio of membrane and fluid
viscosities that determines the degree to which the dissipation is
predominantly two- or three-dimensional \cite{Saffman,Hughes,Lubensky,Ajdari}.
We call this length $\ell_0=\eta_{\rm m}/\eta_{\rm
f}$, where $\eta_{\rm m, f}$ are, respectively, the membrane
(two-dimensional) and fluid (three-dimensional) viscosities. A
similar length, constructed from the two-dimensional shear modulus
and the fluid viscosity determines the scale of deformations
controlled by in-plane versus fluid stresses in the case of
elastic or viscoelastic
membranes\cite{Helfer:2000,Helfer:2001a,Helfer:2001b}.
Here, we consider the motion of extended objects of large aspect
ratio in such quasi-two-dimensional systems of, {\em e.g.}, a
viscous or viscoelastic membrane surrounded by a viscous solvent. In such cases,
both short- and large-scale dissipation can play a role. We
examine in detail the motion of rod-like inclusions, but we
present a general scheme for the calculation of the translational
and rotational mobilities of arbitrary extended bodies.

Even for the case of the motion of simple, rod-like objects in a
bulk Newtonian fluid (\emph{i.e.}, in three dimensions), the situation
is subtle \cite{Lamb,Hughes,Doi}. For instance, for
either infinitely long rods in three dimensions or for the motion
of point-like objects in two dimensions, there is no true low
Reynolds number regime (\emph{i.e.}, linear hydrodynamics) [Batchelor
1967]. Specifically, there is no linear drag coefficient for such a
system: the drag force does not depend linearly with velocity. For
finite length rods and small enough velocities, however, there is
a linear drag coefficient. The drag coefficients for motion
parallel and perpendicular to the rod axis are given by
$$\zeta_\perp=2\zeta_\parallel=4\pi\eta/\ln(AL/a)$$
per unit length, where $L$ is the rod length, $a$ is its radius,
and $A$ is a constant of order unity, depending on the precise geometry.

We examine here a variant of this hydrodynamic
problem, in which the rigid rod lies in a two--dimensional
interface that is viscously coupled to a bulk, Newtonian liquid
phase. The generalization of this problem to the motion of such a rod embedded in
a viscoelastic film is straight-forward. There are many physical
realizations of motion in viscous and viscoelastic films. These
include, for instance, the motion of extended, membrane-associated
proteins in or on the surface of lipid bilayers \cite{Stein,Spooner}. Our work here
is also motivated by recent experiments that have demonstrated the
possibility of constructing and making quantitative measurements
on viscoelastic films that closely resemble cellular structures
such as the {\em actin cortex} \cite{Helfer:2000}. Driven motion of rods in
viscous/viscoelastic films has also been used to determine
rheological properties, e.g., of monolayers \cite{Z:00}. In
addition, the understanding of this problem will allow the
quantitative interpretation of the thermally excited angular
fluctuations of microscopic rod--like particles (such as fd virus)
embedded in such systems. Such hydrodynamic studies may also shed
light on the dynamics of inclusions and transmembrane protein
complexes in fluid cell membranes.  Furthermore, we note that
calculational methods developed in this work allow one to compute
the hydrodynamic drag of irregularly shaped objects embedded in
the film. Such objects can be {\it e.g.\/} fractal aggregates
\cite{Levine:03a}, or lipid rafts in call membranes.

We find two principal results: (i) for small objects
(specifically, for which $L\ll\ell_0$), the drag coefficients
become independent of both the rod orientation and aspect ratio; and (ii)
for larger rods of large aspect ratio, $\zeta_\perp$ becomes
purely linear in the rod length $L$---\emph{i.e.}, the drag becomes
purely local. In contrast, we find that the well-established
three-dimensional result $\zeta_\parallel=2\pi\eta/\ln(AL/a)$
applies for parallel motion in the film, provided that
$L\gg\ell_0$. Here, however, the effective rod radius becomes $\ell_0$ rather than the physical
radius $a$, when $a\ll\ell_0$.
Closely related to (ii), we find that the rotational
drag (equivalently diffusion constant) depends purely
algebraically on the rod length.

We consider only the most simple class of rod geometries in that
we assume the rod to have a circular cross section in the plane
perpendicular to its axis. We take the radius of
that circle to be $a$ while the length of the rod is given by $L$.
The geometry of this rod of length $L$ and radius $a$ is then
parameterized by only one dimensionless ratio, the aspect ratio
$\rho$ defined by
\begin{equation}
\label{aspect-ratio} \rho = \frac{L}{2 a}.
\end{equation}
The long axis of the rod may be assumed to be terminated by
spherical end caps although, as will be seen below, our
calculation is not sensitive to the fine structure of the ends due
to our introduction of a short distance cutoff in the problem. We
will argue below, however, that for the case of large aspect ratio
rods, the detailed structure of the end caps will have a
negligible effect on the overall drag.

The fundamental distinction between the well--known result for the
hydrodynamic drag on a rod in a homogeneous, three--dimensional
viscous material and our result for the drag on a rod embedded in
a viscous membrane coupled to a viscous, three--dimensional fluid
is the appearance of the length scale $\ell_0$. This length enters since the ratio of a
two--dimensional interfacial viscosity $\eta_{\mbox{m}}$ and the
usual viscosity $\eta_{\mbox{f}}$ of the bulk fluid has dimensions
of length,
\begin{equation}
\label{length-scale} \ell_0 = \frac{\eta_{\mbox{m}}}{\sum \eta}.
\end{equation}
The denominator, $\sum \eta$ is the sum of the viscosities of the
fluid above the interface (the superphase) and fluid below the
membrane (the subphase), in the general case of two different
viscosities (or one vanishing one, in the case of Langmuir films).
For a membrane embedded in a uniform fluid of viscosity $\eta$,
this is $\sum\eta=2\eta$. In general, however, we may have two
distinct nonzero viscosities. For a
Langmuir monolayer we have $\sum\eta=\eta$, since we can neglect
the viscosity of air. From here on, however, we shall simply use
the length $\ell_0$ and $\eta_{\mbox{f}}$ for the sum of the
viscosities of the two bulk fluids surrounding the membrane.
Throughout this paper we measure all lengths in terms of the
fundamental length $\ell_0$, except where explicitly stated to the
contrary.

\begin{figure}[htpb]
  \centering
  \includegraphics[width=8.0cm]{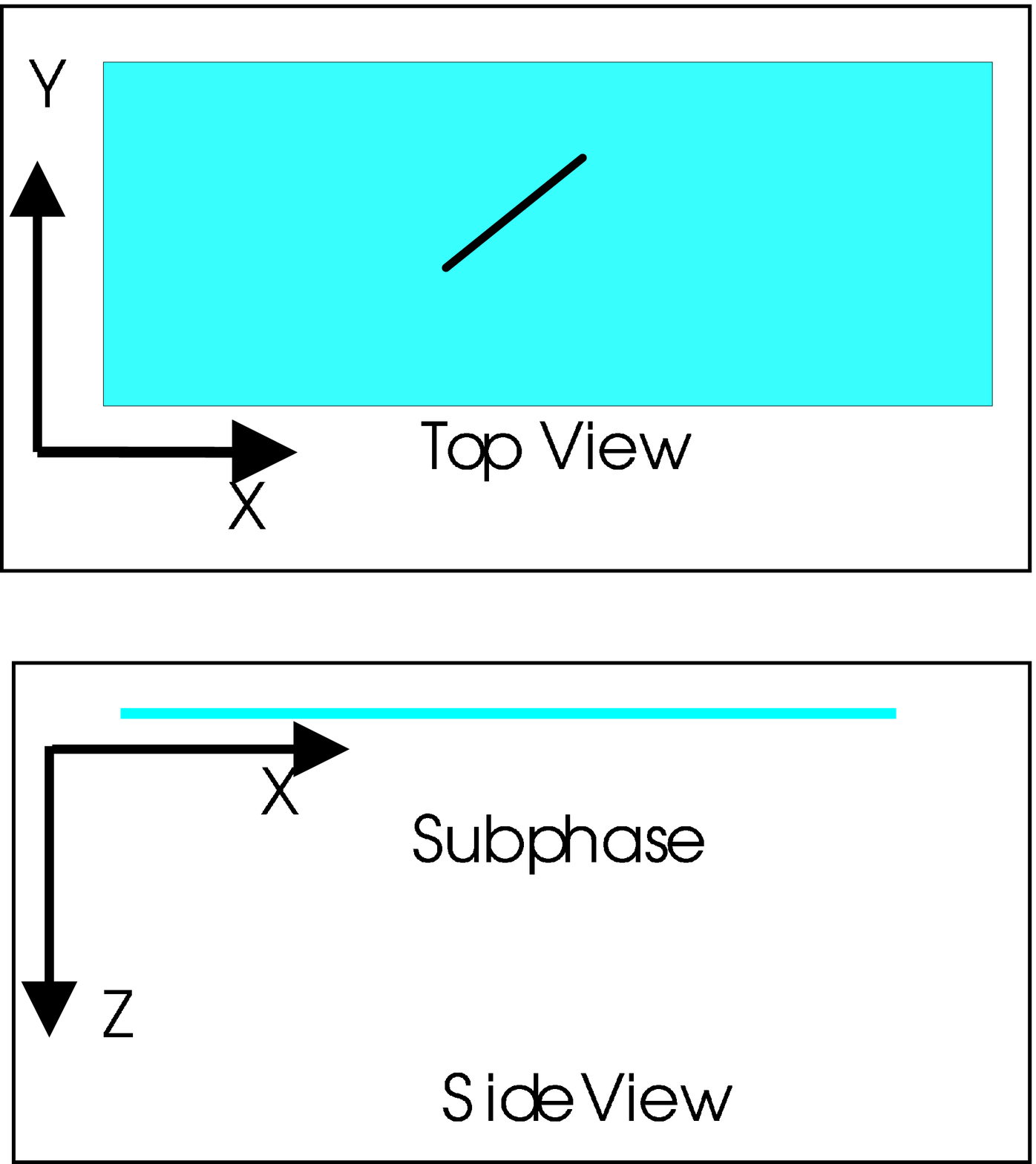}
  \caption{The upper figure shows the film from the top looking down.
The rod is shown as the black line and the film is shown in blue.
The rod is assumed to be embedded in the film, but not in the bulk
subphase below it. This subphase is shown in the lower figure,
which pictures the system from a side view. In this paper the
Newtonian subphase is assumed to be infinitely deep.}
\label{setup}
\end{figure}

There are three independent drag coefficients or particle
mobilities to determine in the problem. The mobility tensor is the
inverse of the drag coefficient tensor. For translational motion,
the mobility tensor $\mu_{ij}$ of the rod in the membrane is
defined by:
\begin{equation}
\label{alpha-def} v^{\mbox{rod}}_{i} = \mu_{ij} F_j,
\end{equation}
where $v^{\mbox{rod}}_{i}$ is the $i^{\rm th}$ component of the
velocity of the rod and $F_j$ is the $j^{\rm th}$ component of the
force applied to the rod.  By in-plane rotational symmetry
combined with the $\hat{n} \longrightarrow - \hat{n}$ symmetry of
the rod, the mobility tensor must take the form:
\begin{equation}
\label{alpha-two-parts} \mu_{ij}= \mu_{||} \hat{n}_i \hat{n}_j
+ \mu_{\perp} \left( \delta_{ij} - \hat{n}_i \hat{n}_j \right).
\end{equation}
Here $\mu_{\perp}$ and $\mu_{\|}$ are the mobility of the rod
dragged perpendicular to and parallel to its long axis
respectively.  In addition to these two independent translational
mobilities, there is also one rotational mobility, $\mu_{\rm
rot}$ linking the angular velocity of the rod to the torque
applied to that rod about its center of inversion symmetry.

It should be noted that there is no hydrodynamic torque acting on
the rod when it is dragged by any force acting through the center
of inversion symmetry of the rod. We refer to these forces as
``central.'' This can be seen from the following argument: Torque
in the two--dimensional plane is a pseudoscalar that must be odd
under time reversal. Such a pseudoscalar can only be built out of
the antisymmetric combination of velocity vector of the rod ${\bf
v}$ and a vector along the axis of the rod ${\bf \hat{n}}$. That
combination $\epsilon_{\alpha \beta} v_\alpha \hat{n}_\beta$ must
also be {\it symmetric\/} under ${\bf \hat{n}}\longrightarrow -
{\bf \hat{n}}$ since the rod is symmetric upon such
transformation. Thus the only available pseudoscalar is
disallowed.  Therefore there is no rotational motion generated by
this class of central forces and the rotational degree of freedom
decouples from the two translational degrees of freedom. We
further note that boundary conditions that break rotational
symmetry or the application of a force at a point other than the
center of symmetry violate the above assumptions and thus allow a
coupling of the rotational and translational motion of the rod.

We approach the solution of this problem by two complimentary
means. In the first part of the calculation, section
\ref{Kirkwood}, we approximate the continuous rod by a series of
discs, in analogy to the Kirkwood approximation used in the
calculation of the drag on a rod in a uniform, three-dimensional
viscous environment \cite{Doi}.  This calculation allows one to
incorporate the details of the shape of the rod in the resulting
drag coefficient. Here the shape of the rod is parameterized in
terms of its dimensionless aspect ratio. This method, however,
becomes numerically difficult in the limit of large aspect ratios,
\emph{i.e.}, for very long, thin rods. In this limit we may
proceed by a second approximation that assumes the rod to be
infinitely thin---\emph{i.e.}, $L \gg a$. Here, we also restrict
our attention to the limit $a \ll \ell_0$.

The outline of the remaining parts of this paper are as follows:
In section \ref{Kirkwood} we develop the first of two
calculational methods for determining the drag on rod--like
objects embedded in the membrane. Then in section
\ref{supermodel-limit} we demonstrate a second approach to
determine the drag on a rod. This approach is optimized to work in
the limit of high aspect ratio rods and compliments the first
approach which is most efficient for rods of small aspect ratio,
\emph{i.e.}, less elongated objects.  The reader who wishes to
examine the results of these calculations without delving into
their details may skip to section \ref{just-the-facts} where the
drag coefficients for translational motion parallel and
perpendicular to the long axis of the rod are computed for a
variety rod geometries. In addition we explore the rotational drag
on these rods there. Finally, we conclude and discuss our
results in section \ref{summary}.

\section{The response function}

\subsection{By the Kirkwood approximation}
\label{Kirkwood}

To incorporate the correct dynamics of this coupled system, we use
the results of our previous calculation\cite{Levine:02} for the
velocity response of the membrane at a position ${\bf x}$ due to
the application of a force localized at the origin. Specializing
to the case of a purely viscous (as opposed to a viscoelastic)
film, we can write the membrane velocity response to a point force
localized at ${\bf x}'$ using the previously defined response
function:
\begin{equation}
\label{def-response}
 v_\alpha \left( {\bf x}\right) =
\alpha_{\alpha \beta}\left({\bf x} - {\bf x}' \right) f_\beta
\left({\bf x}' \right)
\end{equation}
in a closed form as
\begin{eqnarray}
\label{response-tensor}
 \alpha_{\alpha \beta}\left({\bf x}
\right)= &&\alpha_\parallel \left(\left|{\bf x} \right| \right)
\hat{x}_\alpha \hat{x}_\beta +\alpha_\parallel \left(\left| {\bf
x} \right| \right) \left[ \delta_{\alpha \beta} - \hat{x}_\alpha
\hat{x}_\beta \right],
\end{eqnarray}
where the scalar functions $\alpha_\parallel, \alpha_\perp$ of the
distance between the point of the force application and the
measurement of the velocity field are given by:
\begin{eqnarray}
&&-i\omega\alpha_{\|}(x,\omega)=\label{para-def}\\
\nonumber &&{\frac{1}{4\pi\eta_m}}\left[{\frac{\pi}{ z}} {\bf
H}_1(z)-{\frac{2}{ z ^2}}-{\frac{\pi}{2}}
\left(Y_0(z)+Y_2(z)\right)\right]
\end{eqnarray}
and
\begin{eqnarray}
&&-i\omega\alpha_\perp (x,\omega) =\label{perp-def}\\
\nonumber &&{\frac{1}{4\pi\eta_m}} \left[\pi {\bf H}_0(z)-{\frac{\pi}{ z}} {\bf
H}_1(z)+{\frac{2}{ z^2}}-{\frac{\pi}{ 2}}
\left(Y_0(z)-Y_2(z)\right)\right]  ,
\end{eqnarray}
where the ${\bf H}_\nu$ are Struve functions\cite{Abramowitz:64},
and the $Y_\nu$ are Bessel functions of the second kind. We note
$z= \left| {\bf x} \right|/ \ell_0$ is the distance between the
point of force application and the membrane velocity response
measured in the flat membrane in units of $\ell_0$. At
this point, it is important to distinguish between the two
response functions introduced: the mobility tensor of the rod
$\mu_{ij}$ gives the velocity response $v_i$ of the rod given a
total force $F_j$ acting on it through its center. The response
function $\alpha_{ij}({\bf x} - {\bf x}')$ gives the velocity
response $v_i ({\bf x})$ of the two-dimensional {\it fluid\/} at
the position ${\bf x}$ due the application of a point force
$F_j({\bf x}')$ at another point ${\bf x}'$ in the two-dimensional
fluid. The main purpose of this and the next section of the paper
it to derive the former response function for the rod from the
latter response function for the fluid, which we have previously
calculated\cite{Levine:02}.

We note also that, due to the linearity of the underlying
low--Reynolds number hydrodynamics, the velocity response produced
at some point in the membrane by a collection of point forces is
simply the sum of the response functions appropriate to each point
force individually:
\begin{equation}
\label{vel-response}
 v_\alpha \left( {\bf x}\right) =
\sum_{n=1}^N\alpha_{\alpha \beta}\left({\bf x} - {\bf x}_n \right)
f_\beta \left({\bf x}_n \right)
\end{equation}
where $n$ indexes the $N$ point forces located respectively at
locations ${\bf x}_n$ in the film.  Clearly the sum may be
converted into an integral for case of a continuous force
distribution; we will examine this limit for the case of an
infinitely thin rod of finite length in the next section.

\begin{figure}[htpb]
\centering
\includegraphics[width=8.0cm]{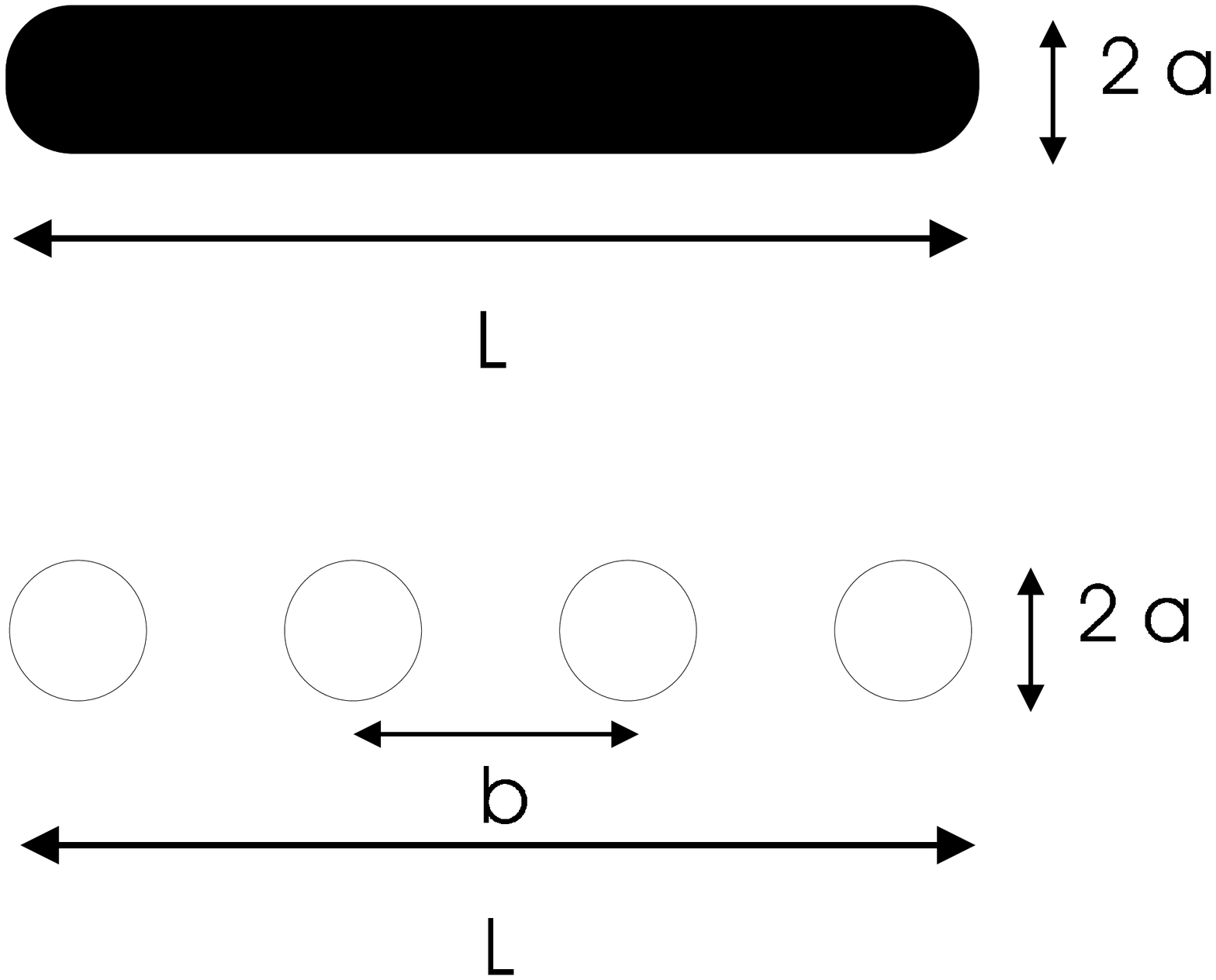}
\caption{Approximating the continuous rod by a series of balls of
radius $a$ with center-to-center separation $b$.  The number of
balls and their separation is chosen so that collection of balls
has the same length $L$ as the original rod. For the best
approximation to the original rod, we maximize the number of balls
for a given aspect ratio of the original rod. }
\label{rodtoballs}
\end{figure}

Using this superposition principle we may determine the effective
drag on a rod by employing the two--dimensional analogue of the
Kirkwood approximation used to calculate the hydrodynamic drag on
a rod in three dimensions. Specifically, we replace the rod of
length $L$ and cross--sectional radius $a$ by a set of $n+1$ disks
of radius $a$ and intersphere separation $b$ chosen so that the
total length of the rod is preserved, {\i.e.\/} $L = nb + 2 a $.
See figure \ref{rodtoballs}. In this way, the aspect ratio of the
rod, $\rho = L/a$ can be fixed. The number of disks, of course,
can be varied, however, we will always choose that number to be
the maximal number consistent with a given aspect ratio and the
non-interpenetrability of the disks. We have also shown that
computed hydrodynamic drag on the rod is only weakly dependent on
the number of disks (or equivalently on the inter-disk
separation).

Our strategy for computing the drag on the rod involves setting
the rod in uniform motion with unit velocity by imposing some set
of forces ${\bf f}^{(i)}$, $i = 1,\ldots,n+1$ on the $n+1$ disks
that make up the rod. Clearly the total force on the rod, which is
equal to the effective drag coefficient is simply the sum of those
forces:
\begin{equation}
\label{force-sum} F_\alpha = \sum_{i=1}^{n+1}f^{(i)}_\alpha =
\zeta_{\alpha \beta} v_\beta
\end{equation}
Using Eq.~(\ref{vel-response}) we can compute the velocity field
for a given collection of point forces. However, to determine the
effective drag on the rod, one needs to insist that all the beads
have the {\it same\/} given velocity and determine the forces
applied to them to ensure this result.

For definiteness, we first discuss the drag on the rod when moving
perpendicularly to its long axis.   We may then directly write out
the analogous solution for motion parallel to the long axis of the
rod.  Motion of the rod along any arbitrary direction in the plane
relative to its long axis can be obtained from these two results
using the linearity of problem.  Thus these solutions span all
possible linear motions of the rod. We will return to rotational
motion shortly.

To calculate the drag on the rod moving perpendicular to its
long axis (see figure \ref{setup}), we first apply
Eq.~(\ref{vel-response})to determine the velocity of all $n+1$
spheres making up the rod as a function of as yet unknown forces.
We insist only, based on the symmetry of this problem, that
these forces are also perpendicular the rod's axis. Thus,
\begin{equation}
\label{drag-perp} v^{(i)} = \sum_{i=1}^{n+1} \alpha_{\perp}^{(ij)}
f^{(j)},
\end{equation}
where we have suppressed the vectorial indices, defined $v^{(i)}$
to be the velocity of the $i^{\mbox{th}}$ sphere at position ${\bf
x}^{(i)}$, and rewritten the mobility tensor using
Eq.~(\ref{perp-def}) and the definition:
\begin{equation}
\label{alpha-tensor}
\alpha_{\perp}^{(ij)}=\alpha_{\perp}\left({\bf x}^{(i)} - {\bf
x}^{(j)} \right).
\end{equation}
The solution for the forces on the individual beads and thus,
using Eq.~(\ref{force-sum}), the total force on the rod and
equivalently the hydrodynamic drag is then obtained by inverting
the $(n+1)\times (n+1)$ matrix of response functions
$\alpha_{\perp}^{(ij)}$. We call this inverse matrix ${\mathcal
M}^{ij}$ defined by:
\begin{equation}
\label{m-def} {\mathcal M}_{\|,\perp}^{(ij)} =
\left(\alpha_{\|,\perp}^{-1}\right)^{(ij)}.
\end{equation}
The drag coefficient is then
\begin{equation}
\zeta_\perp = \sum_{i,j=1}^{n+1} {\mathcal{M}}_\perp^{(ij)},
\end{equation}
where we have used Eq.~(\ref{force-sum}) and the fact that each
bead making up the rod has a velocity of unity. The analogous
expression from the drag on the rod moving parallel to its long
axis is then obtained by replacing the terms of the mobility
matrix, ${\mathcal M}_{\perp}^{(ij)}$, with ${\mathcal
M}_{\parallel}^{(ij)}$ since here all the bead velocities are
parallel to the separation vectors between the beads.  We then
have
\begin{equation}
\zeta_\parallel = \sum_{i,j=1}^{n+1} {\mathcal{M}}_{\|}^{(ij)},
\end{equation}
and the complete solution for motion of the rod in the plane at an
arbitrary angle $\phi$ with respect to its long axis (see figure
\ref{setup}) is given by
\begin{equation}
\label{final-linear-motion} \zeta(\phi) = \zeta_\parallel \cos
\phi + \zeta_\perp \sin \phi.
\end{equation}

It is perhaps not surprising that the drag on the rod does not
have a simple, closed--form solution in the general case. Here we examine the results
of a numerical calculation of the these drag coefficients. The
results depend on two parameters reflecting the relative
magnitudes of three lengths involved in the problem.  We report
our results in terms of the two dimensionless numbers introduced
above: the overall length of the rod (\emph{i.e.}, the linear
dimension of its long axis) measured in units of $\ell_0$, $L/\ell_0$,
and the aspect ratio of the rod $L/a$.

\subsection{Thin rod approximation}
\label{supermodel-limit}

As a calculation tool, the above Kirkwood approximation has one
shortcoming. If one wishes to explore the drag on a very thin rod
compared to its length,{\em i.e.}, one of high aspect ratio, one
is compelled to use a large number of disks. In fact, the number
of disks is linearly proportional to the aspect ratio of the rod
being modelled. Since even numerically inverting a large matrix is
cumbersome, the penultimate step in the procedure outlined above
becomes slow in the high aspect ratio limit. Fortunately, there is
a complementary approach to the Kirkwood scheme that involves
inverting a more manageable matrix and is still valid for
exploring rods of infinite aspect ratio, or specifically, rods
of finite length, but vanishing thickness.  Since it is precisely
in this limit of large aspect ratio that the Kirkwood
approximation becomes intractable, the combination of these two
approaches allows us to study the drag on rod for both large and small aspect ratios.

The thin rod approximation starts from the assumption that one can
take the limit of an infinitely thin rod from the start. Thus the
velocity field at the point ${\bf x}$ due to a continuous
distribution of force densities along the rod, ${\bf f}(x
\hat{x})$ that lies along the $\hat{x}$-axis from $x = -L/2$ to $x
= L/2$ takes the form
\begin{equation}
\label{thin-setup} v_i({\bf x}) = \int_{-L/2}^{L/2} \alpha_{ij}
\left( {\bf x} - p \hat{x} \right) f_j \left( p \hat{x} \right)
dp,
\end{equation}
the indices in the above equation represent the usual vectorial
indices; there are no disks to count here.  There has been a
simplification introduced in Eq.~\ref{thin-setup}. The integral
over the force density, $f \left( {\bf x} \right)$ which in
principal extends over the entire volume of the rod has been made
one-dimensional by assuming the rod to be infinitely thin. This
simplification is possible since the short distance behavior of
the response function includes only an integrable singularity.

After writing Eq.~\ref{thin-setup} one is still faced with the
problem of inverting the equation. After all, the assumed rigidity
of the rod requires that ${\bf v } ({\bf x})$ be constant along
the rod, but force on any length element of the infinitely thin
rod are unknown as they are comprised of both the externally
applied for and, as yet undetermined internal forces of
constraint. Our method of solution is fundamentally identical to
that used in the previous section, we impose a unit velocity field
on the rod and determine the force density required to affect his
result, ${\bf f} \left( {\bf x} \right) $. We use this
intermediate result by integrating the linear force density over
the length of the rod to find the total force necessary to move
the rod with a velocity of unity. This force is clearly just the
drag coefficient that we sought.

The inversion of Eq.~\ref{thin-setup} proceeds by first expanding
the linear force density in Legendre polynomials $P_n(2x/L)$.
From symmetry considerations we can expand the force density in even
Legendre polynomials, \emph{i.e.},
\begin{equation}
\label{force-expansion} f(x) = \sum_{n=0}^N c_{2n} P_{2n}(2x/L)
\end{equation}
for the both the
perpendicular and parallel dragging calculations, while we expand
the force density in odd Legendre functions to study
the rotational drag on the object. In what follows, we only describe in detail the case
of linear drag, although the rotational case is very similar.
The coefficients $c_n$ are as yet unknown, since the function
$f(x)$ is undetermined. In principle since the Legendre
polynomials form a complete set on the interval $-1$ to $+1$, any
physical force density can be expressed as in
Eq.~\ref{force-expansion} provided $N$ be taken to infinity. In
practice we find excellent numerical results even when truncating
this sum to just the first few terms, taking into account the
symmetry of the force distribution about the center of the rod.
We discuss in more detail below
the meaning of numerical accuracy as used in the present context.

Using the expansion of the force density from
Eq.~\ref{force-expansion} we can rewrite Eq.~\ref{thin-setup} as
\begin{equation}
\label{findmatone} {\bf v} \left( {\bf x} \right) = \sum_{n=0}^N
c_{2n} \int_{-L/2}^{L/2} \alpha \left( {\bf x } - z \hat{x} \right)
P_{2n}(2z/L) \, dz.
\end{equation}
To further simplify the analysis, we relax the requirement that
the velocity field be equal to unity at all points along the rod.
Rather we impose this condition at a finite set of points $0\le p_i\le L/2$ along
the rod. Given that we intend to truncate the Legendre function
expansion of the force density at $N$, we can impose the velocity
condition at a maximum of $N+1$ points without creating an
over-determined system of equations. Thus we require that:
\begin{eqnarray}
\label{requirement} {\bf v} \left( p_i \hat{x} \right) &=&0,
{\mbox{for}} \,  i = 0,\ldots,N \\
\label{matequation} 1 &=& \sum_{n=0}^N c_{2n} \int_{-L/2}^{L/2}
\alpha \left( \hat{x} p_i - z \hat{x} \right) P_{2n}(2z/L) \, dz
\nonumber\\
  & &  {\mbox{for}} \, i = 0,\ldots,N.
\end{eqnarray}
Now finding the force distribution along the rod requires only the
inversion of an $(N+1) \times (N+1)$ matrix, ${\mathcal N}_{ij}$ whose
components are defined by
\begin{equation}
\label{matfinal} {\mathcal N}_{ij} =\int_{-L/2}^{L/2} \alpha
\left( \hat{x} p_i - z \hat{x} \right) P_{2j}(2z/L) \, dz.
\end{equation}
Finally, the total force on the rod is found by reconstructing the
force density from its Legendre polynomial expansion and
integrating the resulting expression over the entire rod.  That
force density is given by Eq.~\ref{force-expansion} where the
coefficients, $c_k$ are determined by solving the set of equations
Eqs.~\ref{requirement} using the inverse of the matrix defined in
Eq.~\ref{matfinal}. Thus we find
\begin{equation}
\label{force-solution} c_k = \sum_{i=0}^N {\mathcal N}^{-1}_{ki}.
\end{equation}
Due to the orthonormality of the Legendre polynomials, the total
force is given entirely by the coefficient of the zeroth Legendre
polynomial, $c_0$.  Similarly, the total torque in the case of
rotations is given by the coefficient of the first odd Legendre
polynomial.

There remains one more aspect to this calculation.  How does one
determine the accuracy of the approximation and how can one
optimize that accuracy by selected the set of points along the rod
$x_i, \, i = 0,\ldots,N$ at which to impose the velocity boundary
condition?  We first judge the accuracy of the result by
calculating the velocity field along the rod. Ideally the value of
the velocity should be unity. Because we have only imposed that
boundary condition at a discrete set of points along the rod, the
velocity field along the rod varies. We find that with only eleven
points chosen symmetrically about the center of the rod ({\em
i.e.}, $N=5$ above), the velocity field deviates from unity by at
most five percent in most cases. However, we find that the drag
coefficient (defined by the ratio of total force to average
velocity) converges much more rapidly with $N$. Specifically, we
find a variation of less than a percent in the drag coefficient in
all cases, so long as $N>2$. Below, we report our results for
$N=5$. We thus are able to achieve better than one percent
accuracy with only a $6\times 6$ matrix.

\section{Results}
\label{just-the-facts}

\begin{figure}[htpb]
\centering
\includegraphics[width=8.0cm]{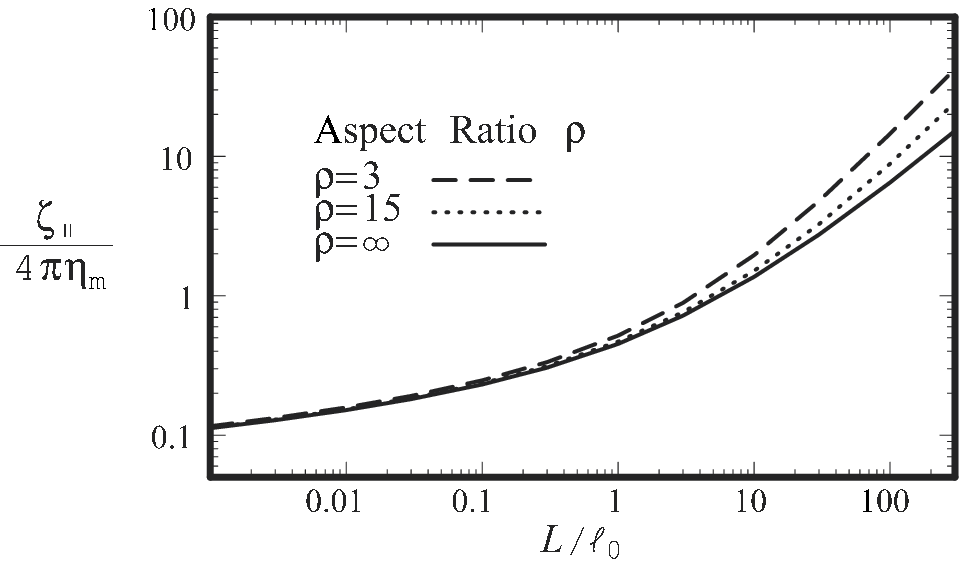}
\caption{Parallel drag coefficient of a rod for various aspect
ratios.} \label{resultone}
\end{figure}

Figure \ref{resultone} shows the parallel drag coefficient for a rod as a
function of its length, $L/\ell_0$ in reduced units.
Two finite aspect ratios, as well as the case of infinite
aspect ratio, {\it i.e.\/}, an infinitely thin rod, are shown in
this figure. As expected based on the above discussion, the shape
of the rod and thus its aspect ratio does not effect the drag on
the rod in the limit all dimensions of the rod are small compared
to $\ell_0$. Essentially all small rod-like particles behave in
the membrane as if they were infinitely thin. For rods longer than
this crossover length, the shape of the rod clearly affects its
hydrodynamic drag as seen by the deviation of the
finite-aspect-ratio curves from the $\rho = \infty$ curve shown as
the solid line in the figure.  It should be noted that the finite
aspect ratio calculations were performed using the Kirkwood
approximation, while the drag on the infinitely thin rod ($\rho =
\infty$) is calculated using the thin-rod approximation.

\begin{figure}[htpb]
\centering
\includegraphics[width=8.0cm]{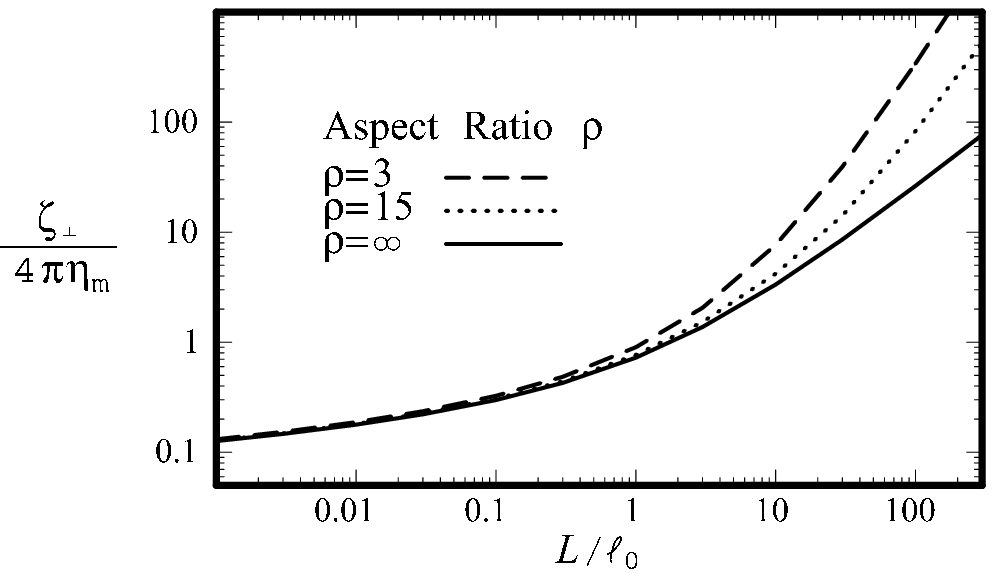}
\caption{Perpendicular drag coefficient of a rod for various
aspect ratios.} \label{resulttwo}
\end{figure}

The analogous calculation can be made for the drag on the rod
moving in a direction perpendicular to its long axis. These
results are shown in figure \ref{resulttwo}.

\begin{figure}[htpb]
\centering
\includegraphics[width=8.0cm]{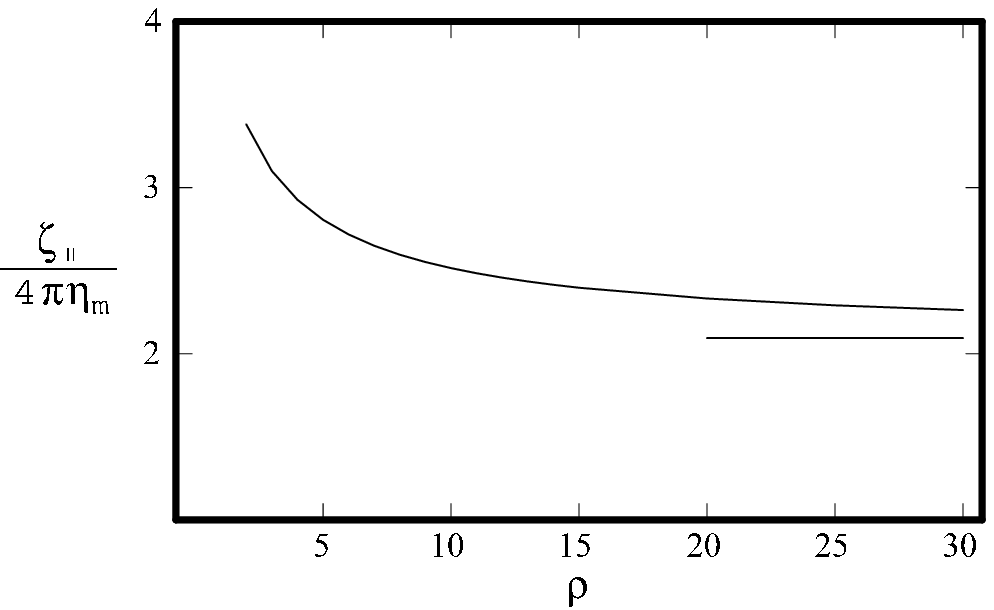}
\caption{The parallel drag coefficient of a finite aspect ratio
rod as a function of aspect ratio, $\rho$. The result for an
infinite aspect ratio is shown as a horizontal line. The length of
the rod is $20 \ell_0$. } \label{resultthree}
\end{figure}

\begin{figure}[htpb]
\centering
\includegraphics[width=8.0cm]{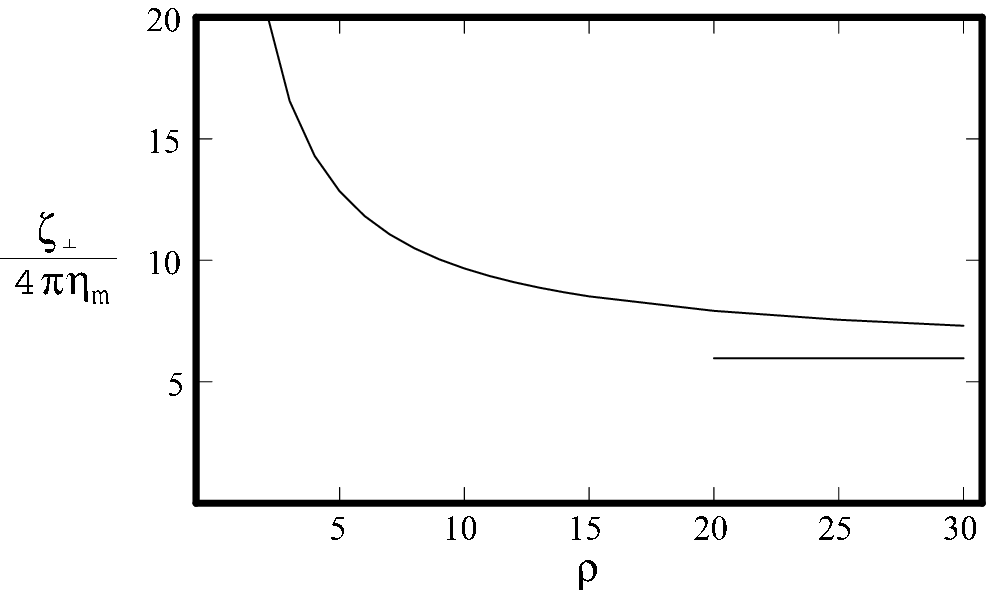}
\caption{The perpendicular drag coefficient of a finite aspect
ratio rod as a function of aspect ratio, $\rho$. The result for
an infinite aspect ratio is shown as a horizontal line. The length
of the rod is $20 \ell_0$ } \label{resultfour}
\end{figure}

The convergence of the finite aspect ratio rod results towards the
infinite aspect ratio result clearly tests the consistency of the
two numerical approaches to drag calculation.  To study in more
detail the aspect ratio dependence of the drag we plot the drag
coefficient for parallel motion (figure \ref{resultthree}) and
perpendicular motion (figure \ref{resultfour}).

\begin{figure}[htpb]
\centering
\includegraphics[width=8.0cm]{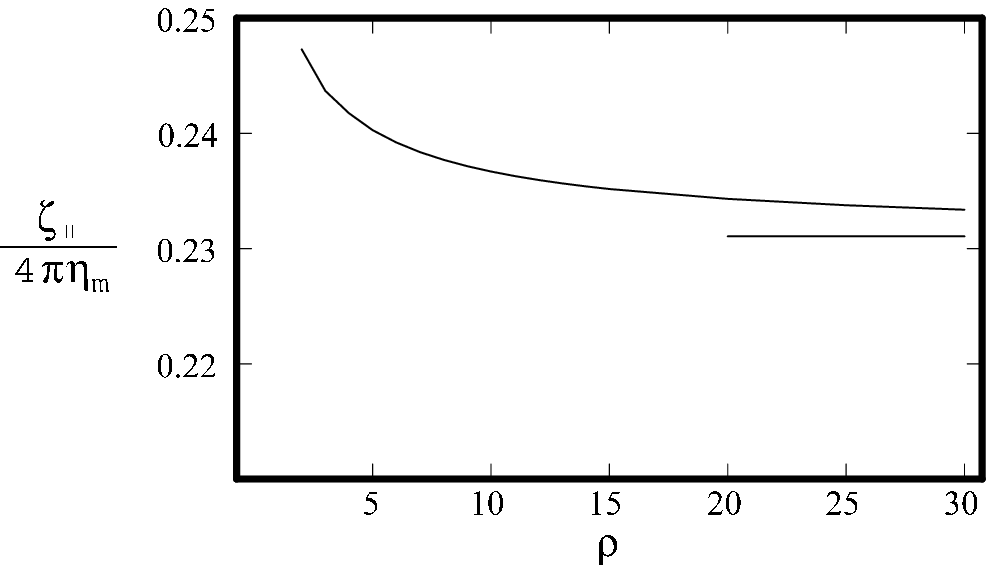}
\caption{The parallel drag coefficient of a finite aspect ratio
rod as a function of aspect ratio, $\rho$. The result for an
infinite aspect ratio is shown as a horizontal line. The length of
the rod is $0.1 \ell_0$ } \label{resultfive}
\end{figure}

\begin{figure}[htpb]
\centering
\includegraphics[width=8.0cm]{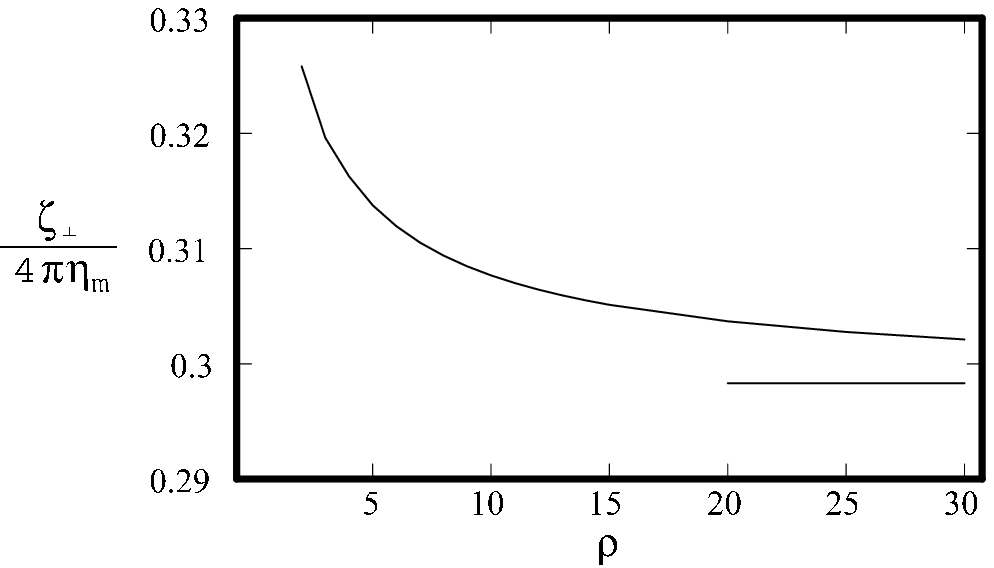}
\caption{The perpendicular drag coefficient of a finite aspect
ratio rod as a function of aspect ratio, $\rho$. The result for
an infinite aspect ratio is shown as a horizontal line. The length
of the rod is $0.1 \ell_0$ } \label{resultsix}
\end{figure}

In both of these figures \ref{resultthree} and \ref{resultfour}
the length of the rod was held constant so that $L/\ell_0 = 20$.
As discussed above particle shape is less relevant for particles
with dimensions less than $\ell_0$ so a rod of significantly
longer length was chosen to explore the aspect ratio dependence of
the rod's drag coefficient. To observe the importance of the
particle size (measured in the natural units of $\ell_0$) we plot
the aspect ratio dependence of the drag coefficient of a rod of
length $0.1 \ell_0$.  The drag coefficient of the rod moving
parallel to its long axis is shown in figure \ref{resultfive},
while the drag coefficient of the rod moving perpendicular to its
long axis is shown in figure \ref{resultsix}.

\begin{figure}[htpb]
\centering
\includegraphics[width=8.0cm]{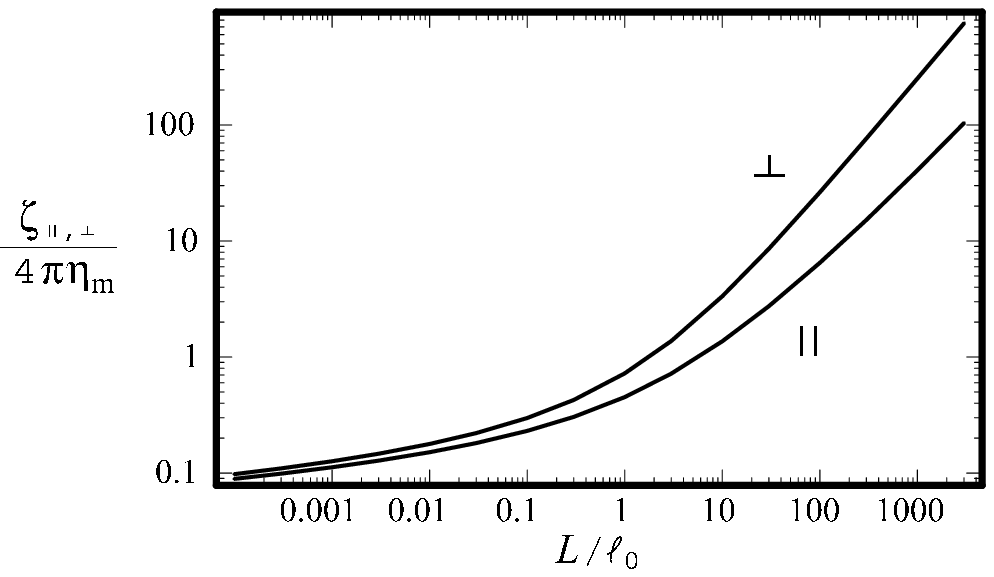}
\caption{ A comparison of the perpendicular and parallel drag
coefficients as a function of rod length for a rod of infinite
aspect ratio. Note that the difference between the two
coefficients is a monotonically increasing function of rod length
and that the two coefficients begin to diverge at the length
$L\simeq \ell_0$.} \label{resultseven}
\end{figure}

It may be observed from a comparison of the pairs of corresponding
figures for parallel drag and perpendicular drag such as
\ref{resultfive} vs. \ref{resultsix}, \ref{resultthree} vs.
\ref{resultfour}, and \ref{resultone} vs. \ref{resulttwo}, that
the perpendicular drag coefficient is strictly larger than the
parallel drag coefficient for rods of all aspect ratios (greater
than one, {\it i.e. not disks\/}) and lengths.  The magnitude of
the  difference between these two drag coefficients, however,
depends on the length of the rod compared with the natural length,
$\ell_0$.  For lengths such that $L < \ell_0$, the two drag
coefficients converge to the same value and the difference
between these coefficients grows monotonically with increasing rod
length. As an example, the two drag coefficients are plotted as a
function of rod length for infinite aspect ratio rods in figure
\ref{resultseven}.

The two coefficients begin to separate at rod lengths on the order
of $\ell_0$.  This differs not only from the small rod limit but also
from the case of motion in bulk fluids, where the two drag coefficients
differ only by a constant factor of two. The length dependence here can be understood by noting
that $\ell_0$ sets the natural length scale over which the
two-dimensional fluid velocity field can vary. In detail, what we find is that the parallel
drag in the film is essentially unchanged from the bulk, three-dimensional drag,
in that
\begin{eqnarray}
\label{parallel-drag} \zeta_{\|} &=& \frac{2 \pi \eta
L}{\ln \left(
0.43 L/\ell_0 \right)},
\end{eqnarray}
where the prefactor in the logarithm has been determined to within $1\%$.
Comparing with the result for drag of a rod in a bulk fluid, the effective radius of the rod is now of order $\ell_0$ (for $a\ll\ell_0$).
On the one hand, that the three-dimensional result is recovered is not
surprising, given that $\ell_0$ corresponds physically to a length scale beyond which
the fluid viscosity dominates the film viscosity. Furthermore, the corresponding fluid velocity
field in this case both respects the assumed incompressibility of the film, and is
the same as that of rod motion in a fluid above this length $\ell_0$. Thus, $\ell_0$
determines the effective aspect ratio.
The case of perpendicular motion, on the other hand, is qualitatively very different. We
find
\begin{eqnarray}
\label{perp-drag} \zeta_{\perp} &=& 2 \pi \eta L.
\end{eqnarray}
Here, the corresponding bulk fluid velocity field in the absence of the film is
inconsistent with incompressibility of the film. Specifically, there is a
non-vanishing two-dimensional divergence of velocity field restricted to the plane of motion
for motion perpendicular to the rod axis. Hence, although only the bulk fluid viscosity $\eta$ enters
this expression (to be expected since dissipation is dominated at the largest scales by the fluid
viscosity), the in-plane incompressibility requires that the fluid velocity field
extends over distances comparable to the largest dimension $L$. This means that the usual
hydrodynamic coupling of portions of the rod (represented by the logarithm) is not present. The result is a
drag coefficient purely linear in rod length. In other words, the drag is effectively \emph{local} in character.

\begin{figure}[htpb]
\centering
\includegraphics[width=8.0cm]{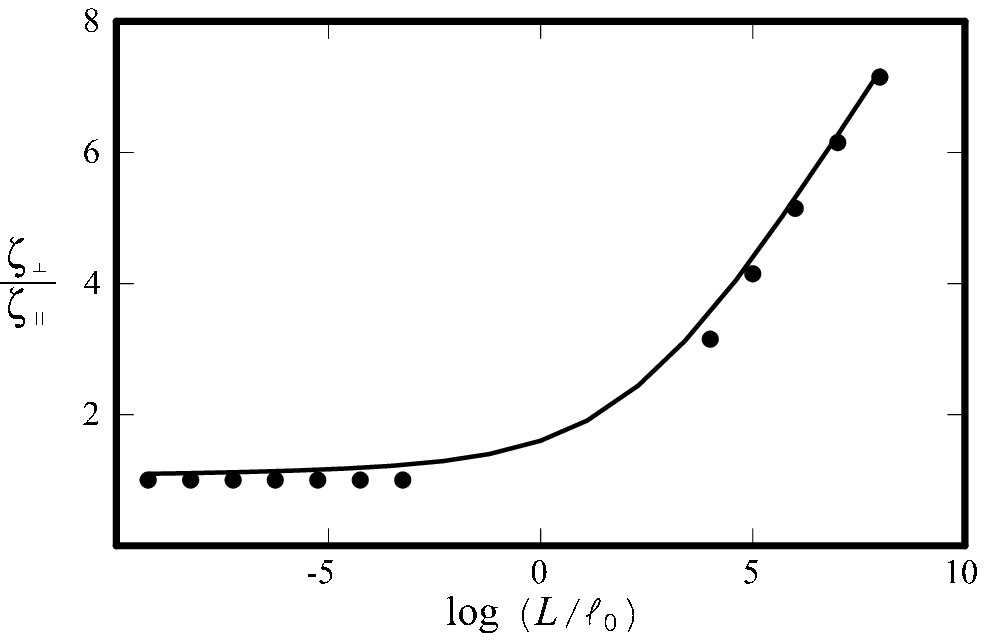}
\caption{The solid line is the ratio of the perpendicular drag
coefficient to the parallel drag coefficient calculated in the
thin rod approximation. The dotted lines are two different
asymptotic fits to this curve corresponding to short rods and long
rods.  For short rods this fit is to the constant one consistent
with the Saffman-Delbr\"uck result. The large rods fit is to a
simple logarithm in length as discussed in the text. }
\label{resulteight}
\end{figure}

\begin{figure}[htpb]
\centering
\includegraphics[width=8.0cm]{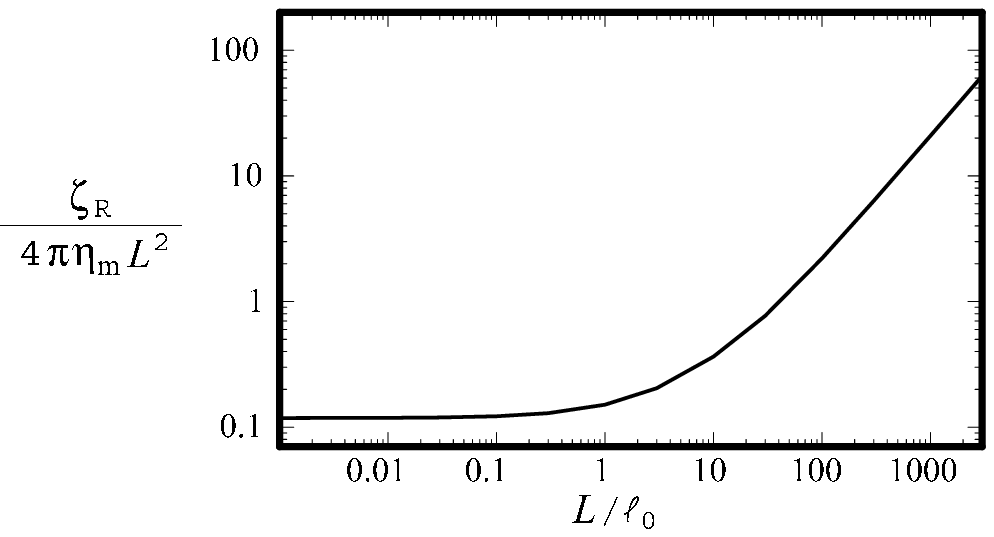}
\caption{The rotational drag coefficient of a rod of infinite
aspect ratio plotted versus the length of the rod. }
\label{resultnine}
\end{figure}

Finally, we note the result for the rotational drag coefficient on the
rod which gives the required torque applied to rod about the
center to generate an angular velocity of the rod equal to unity.
As discussed in the previous sections, this calculation proceeds
analogously to those of the perpendicular and parallel drag
coefficients.  We plot the rotational drag coefficient divided by
$L^2$ for a rod of infinite aspect ratio as a function of the
reduced length in figure \ref{resultnine}. The essential feature
of this plot is that rotational drag coefficient scales as $L^2$
for rods smaller than $\ell_0$ and then as $L^3$ for rods
longer than this natural length. Thus, we find purely algebraic behavior in both limits.

\section{Summary}
\label{summary}

Using the response function previous calculated \cite{Levine:02}
we have calculated the hydrodynamic drag on a rod moving at low
Reynolds number in a viscous film coupled to fluid sub- and superphases of
arbitrary viscosity. The drag coefficient on the rod is a
tensorial object with two independent parameters that correspond
to the drag coefficient of the rod moving along its long axis
(parallel) and in the direction perpendicular to
its long axis. We have also
computed the rotational drag coefficient.

These results were
calculated numerically using two methods with complementary
regimes of validity; the Kirkwood method, which approximates the
rod as a series of non-interpenetrating disks linked together and
is well suited to calculating the drag coefficients for rods of
smaller aspect ratio.  The aspect ratio is set by choosing the
number of these non-interpenetrating disks to make up the rod. For
very long, thin rods having higher aspect ratios, this method
becomes numerically cumbersome since it involves inverting an
$n\times n$ matrix for a rod made up of $n$ disks and reaching
higher aspect ratios requires adding more disks. To explore the
limit of very high aspect ratios, one can perform calculations in the
infinite aspect ratio limit. These two methods can
be shown to be consistent numerically; in the limit of a large
number of disks, we have checked numerically that the results of the
Kirkwood method approach those of the thin rod approximation.

It is instructive
to contrast our results with those for the three--dimensional case. In
three dimensions, there is a length-independent
factor of two difference between the parallel drag coefficient and
the perpendicular drag coefficient:
\begin{eqnarray}
\label{threed-drag} \zeta^{\mbox{3d}}_{\|} &=& \frac{2 \pi \eta
L}{\ln \left(
\frac{A L}{a} \right)} \\
\label{threed-perpar}
 \zeta_\perp^{\mbox{3d}} &=& 2\zeta_{\|}^{\mbox{3d}}.
\end{eqnarray}
The appearance of the logarithm in Eq.~\ref{threed-drag} signals the
break down of a purely local drag (or, ``free draining''). In other words,
the long--range hydrodynamic interactions between various segments of the rod
cause the drag on the rod to be {\em
reduced} from simple linear dependence on $L$ as would be the
case if the hydrodynamic drag on each element of the rod were
purely local in character and thus the total drag additive along
the length of the rod. Instead, the motion of one part of the rod
sets up long-ranged fluid flows that effectively drag other parts
of the rod forward.

The reduced dimensionality of the flow in the film qualitatively
changes this result as show in Eq.~\ref{parallel-drag} and figure
\ref{resultseven}.  From \ref{resulteight} it is clear that the
two drag coefficients are equal in the limit $L \ll \ell_0$ (the
dotted line for small $L/\ell_0$ is simply unity) while in the
limit that $L \gg \ell_0$ they differ substantially. In fact, as
we argue here, we see an apparent, purely local drag per unit
length. Hence, the ratio of the two drags for long rods is given
simply by the logarithm described above, as can be seen by the
asymptotic fit to a logarithm that is illustrated by the dotted
line on the right of the figure.

On the one hand, while the dimensions of the rod are small
($\ll\ell_0$), the dissipation is governed primarily by the film,
which is insensitive to orientation and aspect ratio, just as it
is to size: there is only a weak, logarithmic dependence on size
in this limit \cite{Saffman}. On the other hand, when the rod
becomes longer (than $\ell_0$), the dependence on both orientation
and aspect ratio becomes stronger. Here, the three-dimensional
fluid governs the dissipation, and we know that orientation and
especially size matters in this limit. A major difference,
however, arises when we compare parallel motion with perpendicular
motion. In the latter case, although the dissipation is governed
primarily by the fluid, the film (along with its assumed
incompressibility and no-slip conditions) imposes a very different
boundary condition on the flow from what we would have in a bulk
fluid alone. The velocity field $\vec v_\perp$ in the film
satisfies $\vec\nabla_\perp\cdot\vec v_\perp=0$, which is
inconsistent with the Stokes flow in perpendicular motion.
Considered as a two-dimensional field, $\partial_x v_x+\partial_y
v_y$ is non-zero in the plane of motion in this case. This added
condition can only increase the drag for perpendicular motion
relative to that without the film present. In contrast, since this
boundary condition {\em is} consistent with the flow field for
parallel motion, we expect to find quantitative agreement with the
parallel mobility in a bulk fluid when the film's viscosity
becomes irrelevant (small $\ell_0$).

For the perpendicular motion, not only is the drag increased relative to that for a
bulk fluid, but the dependence is purely linear, as show in Fig.\ \ref{resultseven}.
The linear dependence is an indication of the absence of the
hydrodynamic effects described above, which reduce the drag by
cooperativity of sections along the rod. Here, the drag is simply
proportional to the length of the rod. This can be seen from the
additional boundary condition mentioned above. Although the
dissipation for long rods is governed by the fluid viscosity, the
characteristic scale for this flow is that of the whole rod.
Unlike the case of perpendicular motion in a simple fluid, where
there is a short path of order the rod diameter $a$ around the
rod, the in-plane incompressibility forces the flows to go around
the long way. Hence, the total absence of the logarithm, and the
simple, purely local drag proportional to length. This can be seen
in Fig.\ \ref{resulteight}, where we show the ratio of the drag
coefficients is just given by the logarithmic term coming from the
hydrodynamics of rods in ordinary fluids. Finally, we note that
these observations also explain why the rotational drag is purely
algebraic for long rods, since rotations exhibit a purely local
drag or rod segments perpendicular to the motion. Specifically, as
shown in Fig.\ \ref{resultnine}, we find
\begin{eqnarray}
\label{rot-drag} \zeta_{R} &\simeq& 0.16 \pi \eta L^3.
\end{eqnarray}

In summary we have developed a highly adaptable framework to
compute the drag on irregularly shaped objects embedded in a
viscous membrane or interface.  As a demonstration of this method
we have computed the drag on a rigid rod in this two-dimensional
fluid system viscously coupled to a fluid subphase and compared
our results to both the well-known results for the drag on a rod
in a three-dimensional, viscous fluid, and the result for the drag
on a disk embedded in a membrane (due to Saffman and Delbr\"uck),
which is applicable to {\it e.g.\/} the diffusion of small
transmembrane proteins.  Firstly, we find, in accordance with the
results of previous investigators \cite{Hughes,Lubensky,Ajdari}
that there is an inherent length scale $\ell$ in the system set by
the ratio of the two-dimensional viscosity of the membrane to the
three dimensional viscosity of the subphase fluid.  The existence
of such a length scale is made obvious by dimensional analysis.
The importance of this length scale on the drag tensor associated
with various objects embedded in the membrane has been discussed
in this work. In brief, for objects with characteristic dimensions
less than $\ell$, the Saffman-Delbr\"uck result is recovered from
our more general computation. The drag coefficient tensor for all
such small objects is simply that of a small disk. It is isotropic
and independent of particle size except for logarithmically small
corrections. For objects significantly larger than $\ell$, our
analysis shows that drag tensor becomes both size dependent and
anisotropic for rod-like objects. Based on our calculations
restricted to rod--like objects, it is nevertheless clear that
these general statements apply to objects of arbitrary shape as
well. In particular for these rods, we find that the parallel drag
coefficient matches that of a rod dragged parallel to its long
axis in a three-dimensional, viscous fluid.  For the same rod
dragged perpendicular to its length a qualitatively different
result is found: the effective drag coefficient is logarithmically
enhanced versus its three-dimensional counter-part due to the
breakdown of long-range hydrodynamic interactions in the interface
due to momentum transfer to the bulk, subphase.

While the heuristic importance of this new length scale is clear,
we have not yet estimated its size for typical membranes. As a
starting point, we note that if the membrane/interfacial viscous
were equal to that of the bulk, subphase, this length would
naturally be the thickness of the membrane, {\it i.e.\/} a
molecular length. For a short chain surfactant monolayer or lipid
bilayer this length would then be on the order of $1 - 2$nm
respectively.  However, it is expected that the internal viscosity
of the membrane/interface is typically much larger than that of
the (typically aqueous) subphase and this length $\ell$ is
consequently multiplied by a factor equal to this viscosity
enhancement of the membrane/interfacial material. It is not
unreasonable to suppose that $\ell \sim 10 - 100$nm. Thus we
expect to see significant deviations from the Saffman-Delbr\"uck
result both for lipid rafts and protein aggregates that surpass
such lengths. At the same time, it is clear that the standard
Saffman-Delbr\"uck result should explain the observed mobility of
individual transmembrane proteins.

Further experimental tests of the above theory require the
analysis of tracer particle diffusion data for a
membrane/interface bound objects of various sizes. Based on these
calculations, the observation of anisotropic diffusion constants
for these longer rod-like objects would be a clear indication of
phenomena unexplainable by the Saffman-Delbr\"uck analysis. After
testing these basic mobility calculations, one could then use
these results to do both standard translational microrheology on
membranes and interfaces\cite{Levine:02} as well as novel
experiments on rotational microrheology\cite{Mason}. Such studies
will be particularly interesting in studying the properties of
broken rotational phases of lipid monolayers such as hexatic
phases\cite{Dennin}.  In addition one should be able to extend the
present calculations to explore the implications of membrane
hydrodynamics upon diffusion limited aggregation in membranes.
Such aggregation processes play an interesting role in the
formation of transmembrane protein aggregates, lipid rafts, and
colloidal aggregates on large, unilamellar
vesicles\cite{Dinsmore}.

\acknowledgments

AJL would like to thank C.~Alonso and other members of the
Zasadzinski group for frequent discussions. The authors
acknowledge the hospitality of the Kavli Institute for Theoretical
Physics where most of this work was performed. In addition AJL
would like to acknowledge the hospitality of the Vrije
Universiteit, Amsterdam. Finally, the authors would especially
like to thank D.K.~Lubensky for helpful conversations on this
problem.  This work is supported in part by the National Science
Foundation under Grant Nos. DMR98-70785 and PHY99-07949.

\end{document}